\documentclass[]{article}

\usepackage[english]{babel}
\usepackage{amsmath, amsthm, amsfonts, amssymb, accents}
\usepackage{graphicx, dsfont, srcltx, color}

\usepackage{lineno,hyperref}

\theoremstyle{plain}
\newtheorem{theorem}{Theorem}

\def\Tht{\Theta}
\def\tht{\theta}

\def\l{\lambda}

\def\D{\Delta}

\def\b{\beta}

\def\di{\,d}

\def\d{\delta}

\def\Op{\mathcal{H}}
\def\H{H}

\newcommand{\PT}{\mathcal{PT}}

\DeclareMathOperator{\IM}{Im}

\begin{document}


\title{Sequences of closely spaced resonances and eigenvalues for  bipartite complex potentials}
\author{D. I. Borisov$^{1,2,3}$
$^1$Institute of Mathematics, Ufa Federal Research Center,
\\
Russian Academy of Sciences,
Ufa, Russia, 450008
\\[3mm]
$^2${Bashkir State Pedagogical University named after M.Akhmulla}, 
Ufa, Russia, 450000
\\[3mm]
$^3$University of Hradec Kr\'alov\'e,
Hradec Kr\'alov\'e 50003, Czech Republic
\\[5mm]	
D. A. Zezyulin$^4$\\
$^4$ITMO University, St. Petersburg 197101, Russia}

\date{\empty}

\maketitle

\begin{abstract}
We consider a  Schr\"odinger operator on the axis with a bipartite potential consisting of  two compactly supported complex-valued functions, whose supports are separated by a large distance.  We show that this operator possesses a sequence of approximately equidistant  complex-valued wavenumbers   situated near  the real axis. Depending on its imaginary part, each wavenumber corresponds to either a resonance or an eigenvalue.
The obtained  sequence of wavenumbers  resembles   transmission resonances in electromagnetic Fabry-P\'erot interferometers formed by parallel mirrors.
Our result   has potential applications in standard and non-hermitian quantum mechanics,   physics of waveguides, photonics,  and in other areas where the  Schr\"odinger operator  emerges as an effective Hamiltonian.
\end{abstract}




\section{Introduction and motivation}
\label{sec:model}

Resonances play an important role in quantum mechanics \cite{Taylor} and in other branches of physics, where the interaction between a  wave and a  localized potential is described by an effective Schr\"odinger Hamiltonian. While the standard quantum mechanics deals with real-valued potentials, there is a steadily growing interest in non-hermitian quantum theories  which involve complex potentials \cite{Muga,Bender,Most,Moiseyev}. Apart from this, non-self-adjoint Schr\"odinger Hamiltonians is a hot topic in many  other fields,  including theory of electromagnetic waveguides with gain and absorption  \cite{Mostafazadeh2009}, non-hermitian photonics  \cite{FengReview}, and parity-time-symmetric physics \cite{KZYreview}.

In this Letter, we consider the resonances and eigenvalues of a one-dimensional Schr\"odinger operator with  a complex potential formed  by  two 
functions with compact supports separated by a sufficiently large distance. Let $V_\pm=V_\pm(x)$, $x\in\mathds{R}$, be two measurable bounded complex-valued functions and $\ell>0$ be a   parameter.  Suppose that the function $V_+$ is supported in $[0,d_+]$,   while $V_-$ is supported in $[-d_-,0]$,  $d_\pm>0$, and  introduce the Schr\"odinger operator
\begin{equation*}
\Op_\ell:=-\frac{d^2\ }{dx^2} + V_\ell(x),\qquad V_\ell(x):=V_+(x-\ell) + V_-(x+\ell),
\end{equation*}
on $\mathds{R}$. Rigorously we define $\Op_\ell$ as an unbounded operator in $L_2(\mathds{R})$ on the domain $\H^2(\mathds{R})$. This is an $m$-sectorial operator; if $V_\pm$ are real-valued functions, this operator is self-adjoint.

The main object of our study is the resonances and the eigenvalues of the operator $\Op_\ell$.  They correspond to
$k\in\mathds{C}$, for which the problem
\begin{equation}\label{2.2}
-\psi''+V_\ell(x)\psi=k^2\psi,\qquad x\in\mathds{R},
\qquad
\psi(x)=C_\pm e^{{\pm} i k x},\qquad \pm x>x_0, \qquad C_\pm= \textrm{const},
\end{equation}
possesses a non-trivial solution. Here $x_0$ is fixed number such that the function $V_\ell$ vanishes outside $[-x_0,x_0]$, for instance, $x_0=\ell+d_-+d_+$, and the prime denotes the derivative with respect to $x$. Wavenumbers  $k$ with {$\IM k \leqslant 0$ describe 
the resonances $\l=k^2$} associated with   non-$L_2$-integrable eigenfunctions $\psi$. Wavenumbers $k$ with {$\IM k > 0$} correspond to {the} eigenvalues   $\l=k^2$ associated with integrable bound states $\psi$ having a finite $L_2$-norm.

\section{Main result}

Our main result states  that as the distance $\ell$ is sufficiently large, problem (\ref{2.2})  features a sequence of approximately equidistant values $k=k_n$ corresponding to resonances or eigenvalues. 
Proceeding  to the rigorous formulation of the main result,  we introduce some additional notations.
By $X_\pm$, $Y_\pm$  we denote the
solutions of the problems
\begin{equation}
\begin{aligned}
-&Y_\pm'' + V_\pm Y_\pm=k^2 Y_\pm,\qquad  && x\in\mathds{R}, &  -&X_\pm'' + V_\pm X_\pm=k^2 X_\pm,\qquad  &&  x\in\mathds{R},
\\
&  Y_\pm(x,k)=e^{i k x}, && x>d_+,\qquad  && X_\pm(x,k)=e^{-i k x}, && x<-d_-.
\end{aligned}\label{2.3}
\end{equation}
Next,  we introduce  a function
{
\begin{equation*}
F(k):=\frac{X_-'(0,k)-ik X_-(0,k)} {X_-'(0,k)+ik X_-(0,k)} \frac{X_+'(d_+,k)-ik X_+(d_+,k)}{X_+'(d_+,-k)-ik X_+(d_+,-k)}.
\end{equation*}}
By $\circ$ we denote the superposition of the functions, that is, $(f\circ g)(k)=f(g(k))$. An $m$-multiple superposition of a function $f$ is denoted by $f^{[m]}$, i.e., $f^{[m]}=\underbrace{f\circ f\circ\cdots\circ f}_{m\ \text{times}}$.  The symbol $\lfloor\cdot\rfloor$ stands for the integer part of a number.

Our main result is the following theorem. Its proof is presented below in  {S}ection~\ref{sec:proof}.

\begin{theorem}\label{th1}
Assume that
\begin{equation}\label{2.4}
X_-'(0,0)\ne0,\qquad X_+'(d_+,0)\ne0.
\end{equation}
There exists $r>0$ depending on $V_\pm$ only, such that the function $F(k)$ is well-defined, holomorphic and non-zero on $B:=\{k\in\mathds{C}:\, |k|\leqslant r\}$. For all integer $n\in\mathds{Z}$ obeying $|n|\leqslant N_\ell$,
$N_\ell:=\left\lfloor \frac{2\ell r}{\pi}-\frac{1}{2}\right\rfloor$,
the circle $ B_n:=\big\{k\in\mathds{C}:\, |k-a_n|<\tfrac{\pi}{4\ell}\big\}$,
\begin{equation}\label{an}
a_n:=\frac{\pi n}{2\ell},
\end{equation}
possesses exactly one value $k_n(\ell)$, for which problem (\ref{2.2}) has a non-trivial solution. As
\begin{equation}\label{2.8}
\frac{e^\frac{\pi}{2}}{4\ell}\max\limits_{\overline{ B}} |F'|<1,
\end{equation}
the value $k_n$ satisfies the representation
\begin{equation}\label{2.9}
k_n=a_n+\lim\limits_{m\to+\infty} h_n^{[m]}(0), \qquad h_n(k):=-\frac{i}{4\ell}\ln F(k+a_n),\qquad
\big|k_n-a_n-h_n^{[m]}(0)\big|\leqslant \frac{\pi e^{\frac{m\pi}{2}}}{(4\ell)^{m+1}}  {\max\limits_{\overline{B}}}^m |F'|,
\end{equation}
where the branch of the logarithm is fixed by the condition $\arg\ln z\in(-\pi,\pi]$ and $m\in\mathds{N}$. The value $k_n$ 
can be also represented by   absolutely uniformly in $\ell^{-1}$ convergent series
\begin{equation}\label{2.11}
k_n=a_n + \sum\limits_{m=1}^{\infty} \frac{(-i)^m}{4^m m!\ell^m} \frac{d^{m-1}\ln^m F}{dk^{m-1}}\bigg|_{k=a_n}=\sum\limits_{m=1}^{\infty} \frac{1}{2^m m! \ell^m} \frac{d^{m-1}\ }{dk^{m-1}}\left(\pi n - \frac{i}{2}\ln F(k)\right)^m\bigg|_{k=0}.
\end{equation}
For all $M\in\mathds{N}$ the inequalities hold:
\begin{align}\label{2.12}
	&\left|k_n-a_n - \sum\limits_{m=1}^{M} \frac{(-i)^m}{4^m m!\ell^m}  \frac{d^{m-1}\ln^m F}{dk^{m-1}}\bigg|_{k=a_n}\right|\leqslant \frac{1}{4^{M+1}(M+1)!\ell^{M+1}} \max\limits_{\overline{B}} \left|\frac{d^M\ln^{M+1} F}{dk^M}\right|,
\\
&\left|k_n-\sum\limits_{m=1}^{M} \frac{1}{2^m m! \ell^m} \frac{d^{m-1}\ }{dk^{m-1}}\left(\pi n - \frac{i}{2}\ln F\right)^m\bigg|_{k=0}\right|\leqslant \frac{1}{2^{M+1}(M+1)!\ell^{M+1}} \max\limits_{\overline{B}} \left|\frac{d^M\ }{dk^M}
\left(\pi n - \frac{i}{2}\ln F\right)^{M+1}
\right|.\label{2.12a}
	\end{align}
\end{theorem}

\section{Discussion of the main result}


Our main result states that as $\ell$ is not too small so that $N_\ell\geqslant 1$, there exist at least $2 N_\ell+1$ complex-valued wavenumbers
 $k_n$, $|n|\leqslant N_\ell$, such that each $k_n$ corresponds to  a nontrivial solution of problem   (\ref{2.2}) and therefore represents either a resonance or an eigenvalue of the operator $\Op_\ell$. Values  $k_n$  are located in small circles $B_n$ centered at the points
$a_n$ in (\ref{an}),  one  value  in each circle,  and  form 
an approximately equidistant sequence situated close to the real axis. As $\ell$ increases, the number of  the  eigenvalues and resonances
grows proportionally to $\ell$, while the distances between neighbouring    wavenumbers $k_n$ 
tend to zero.

The obtained result is fairly general and holds independently of the specific shape of left ($V_-$) and right ($V_+$) components of the potential $V_\ell$, provided that $V_\pm$ have finite support and are separated well enough. Potential $V_\ell$ may be real or complex-valued, and therefore each   $k_n$   represents either a  resonance or an isolated (generically, complex) eigenvalue  of non-self-adjoint operator  $\Op_\ell$.  Nonzero values $k_n$ that eventually lie on the real axis  correspond to spectral singularities (i.e., zero-width resonances). An important feature of our result is that it is valid not only for potentials $V_\pm$, but also for more general perturbations. Namely, we can replace the operators of multiplications by $V_\pm$ by more general operators but still acting on $[-d_-,0]$ and $[0,d_+]$, and the statement of Theorem~\ref{th1} remains the same. The main result is formulated only in terms of the function $F$   which is determined by   the functions $X_\pm$ and their derivatives at certain points; no other information on $V_\pm$ is needed. This is why, for instance, we can assume that $V_\pm$ describe  second-order differential operators, i.e.
$V_\pm u=A_\pm^{(2)}u'' +A_\pm^{(1)}u' +A_\pm^{(0)}u$, where $A_\pm^{(j)}$ are  smooth compactly supported  on $[-d_-,0]$ and $[0,d_+]$ complex   functions. Another generalization for which Theorem~\ref{th1} holds after some obvious modifications
corresponds to a pair of delta-interactions  $V_\pm=\b_\pm \d(x)$, where $\b_\pm$ are   complex constants (see Section~\ref{sec:examples}).

A similar spectral picture was described in the recent study \cite{SW1} devoted to  the resonances of a one-dimensional discrete Schr\"odinger operator. The potential was of the form $V\chi_L$, where $V$ was either periodic or random, and $\chi_L$ was a characteristic function of the interval $[-L+1,L]$. It was shown, that as $L\to+\infty$, there was a sequence of closely spaced resonances accumulated along some analytic curve for the periodic potential or even a cloud of closely spaced resonances located in some domain for the random potential. Here the points $-L+1$ and $L$, at which the potential $V$ was replaced by zero, played the same role as our potentials $V_\pm$, and this explains the similarity of our spectral picture with that in \cite{SW1}. The main result in \cite{SW1} for  periodic $V$ provided   the existence of the resonances, the leading terms of their asymptotics and asymptotic description of the distribution density  of the resonances. The  same continuous model for the Schr\"odinger operator was treated in \cite{BG}. It was shown that there existed a sequence of closely spaced of resonances accumulated along a certain curve.
We   also note that  in \cite[Ch. I\!I.1, Sect. I\!I.1.4, Thm. 1.4.1]{Al}, the operator $-\D+ V_1+V_2(\cdot-\ell y)$ in $\mathds{R}^3$ was considered, where $V_{1,2}$ were real compactly supported functions and $y$ was a fixed point. It was proved that this operator possessed an infinite sequence of resonances converging to zero as $\ell\to \infty$, and their asymptotic behaviors were $k_n=\ell^{-1} k_{0,n}+\ell^{-2} k_{1,n} + o(\ell^{-2})$, where $i k_{0,n}|y|$ were the roots of the equation $z=\pm e^{z}$. Since the error term in this asymptotic law was non-uniform in $n$, this expansion was useful only for sufficiently large $\ell$ and some fixed $n$ and could   not be effectively applied as $n\sim\ell$. No other information on the resonances was obtained in the cited book, so, this result just indicated the existence of some sequence of resonances without careful studying  its properties.
As we shall discuss below, our result provides much more information about the location of the resonances and, moreover, in our case there can be also ladders of eigenvalues or mixed ladders of coexisting eigenvalues and  resonances since our potentials $V_\pm$ are not assume to be real-valued.

In our model, the sequence  of eigenvalues and resonances emerges due to the  large distance
between the  supports of $V_\pm$. The Schr\"odinger operators with several potentials separated by a large distance is a classical subject \cite{D, H, KS1, UMJ}. The results of \cite{UMJ} applied to our operator $\Op_\ell$  state that as $\ell\to+\infty$, the resolvent $(\Op_\ell-\l)^{-1}$ splits into a direct sum of three resolvents $(\Op_\pm-\l)^{-1}$ and $(\Op_0-\l)^{-1}$, where
$
\Op_\pm:=-\frac{d^2\ }{dx^2}+V_\pm$, $\Op_0:=-\frac{d^2\ }{dx^2}$.
The spectrum of $\Op_\ell$ converges to the union of the spectra of $\Op_\pm$ and $\Op_0$. The essential spectrum of each of these operators is $[0,+\infty)$ and contains \textsl{no} embedded eigenvalues. In the vicinity of zero, the operators $\Op_\pm$ and $\Op_0$ can have only finitely many resonances. In other words,
the operators $\Op_\pm$ and $\Op_0$ have only finitely many singularities of a (meromorphic continuation of) the resolvent. Despite this fact, as the distance between the  supports of $V_\pm$ is large enough,  the sequence of closely spaced eigenvalues and resonances of the operator $\Op_\ell$ emerges.

Turning to the physical interpretation of our effect, it is relevant to point out that  the constructed sequence of wave numbers resembles    resonances emerging  in optical Fabry-P\'erot interferometers formed  by two parallel plates \cite{Born}.   If   light is launched in such a resonator under normal incidence, then the constructive    interference between internal  reflections of light   traveling between the plates results in a sequence of transmission resonances at wave vectors $k_n^{(FP)} = n\pi/L^{(FP)}$, where  $L^{(FP)}$ is the geometrical length of the resonator. Considering the   components of the bipartite potential  as the ``plates'' of the resonator and noting that for large $\ell \gg 1$ the distance between the ``plates'' can be interpreted as the length of the resonator, i.e.,  $L^{(FP)} = 2\ell$, we observe that the obtained expressions for the Fabry-P\'erot resonances $k_n^{(FP)} = (n\pi)/(2\ell)$  coincide with    the ball centers $a_n$ in (\ref{an}) that give the location of our wavenumbers  in the leading order. Therefore the obtained result can be interpreted as a Fabry-P\'erot interferometer for waves governed by  by effective Schr\"odinger-type Hamiltonians. This result extends  the well-known analogy between the Fabry-P\'erot interferometer and  resonances of a quantum particle scattered   by a single potential well or a barrier \cite{Cohen}.

Our sequences of resonances and eigenvalues could also be a reflection of some kind of tunneling between the potentials $V_\pm$. At the same time, usually, the tunneling leads to an exponentially small asymptotic law for the eigenvalues, even in a very general case \cite{GAM2}, while our resonances and eigenvalues exhibits power-law asymptotic behavior, see (\ref{2.11}). The exponential asymptotic laws for usual tunneling are due to the exponential fall-off of the eigenfunctions for each single potential, while for the resonances the situation can be different and one can face an asymptotic power law. However, our sequence can also include eigenvalues.

Apart of the existence of the resonance and eigenvalues,
Theorem~\ref{th1} also provides two recipes for finding explicitly the corresponding wavenumbers $k_n(\ell)$.  The first of them is  the formula in (\ref{2.9})  defining a convergent  iterative process  $k_n^{(m)} := a_n + \eta^{(m)}_n$, where $m\in \mathds{N}$ counts the iterations, and $\eta^{(m)}_n$ is obtained recurrently as    $\eta^{(m+1)}_n := (4i\ell)^{-1}\ln F(\eta^{(m)}_n + a_n)$, where $\eta^{(1)}_n := (4i\ell)^{-1}\ln F(a_n)$.  This approach is perfectly suited  for the numerical finding of $k_n$. The estimate in  (\ref{2.9})  controls the error between the exact value  $k_n$ and the   approximation  $k_n^{(m)}$   after $m$ iterations; this estimate is \textsl{independent} of $n$.
The radius $r$ of the circle $B$ is some implicit characteristics of the potentials $V_\pm$. As we shall show in the proof of Theorem~\ref{th1}, this radius must be chosen such that  $|F(k)-1|<   1-e^{-\frac{\pi}{2}}\approx 0.792$ as  $|k|\leqslant r$.

Another approach to finding $k_n$ is provided by two absolutely uniformly convergent series (\ref{2.11})  which should be regarded as   Taylor series for $k_n$ in powers of $-\frac{i}{4\ell}$ either at zero or at the point $a_n$, no matter that $a_n$ depends on $\ell$ as well. The partial sums of these series  approximate $k_n$ according to  (\ref{2.12}) and (\ref{2.12a}).
We stress that the right-hand side of the inequality (\ref{2.12}) is independent of $n$.   The second series in equation  (\ref{2.11}) is well-adapted for providing asymptotic behavior for $k_n$ as $\ell\to\infty$.
Despite now the coefficients  are independent of $a_n$,
the price we have to pay is a worse error term in (\ref{2.12a}): now it depends on $n$ and as $n\sim \ell$, the error term is of order constant for all $M$. Nevertheless, the second series in (\ref{2.11}) describes very well and in an explicit form the behavior of $k_n$ for large $\ell$ and not very large $n$. We also observe that the leading terms in the first series in (\ref{2.11}) are $k_n\cong a_n-i(4\ell)^{-1}\ln F(a_n)$ and hence, the real part of the function $\ln F(k)$ determines the sign of the imaginary part of $k_n$. As $k$ ranges in $[-r,r]$, the real part of $\ln F(k)$ can change the sign and this is why, our {sequence} can contain simultaneously eigenvalues and resonances.

\section{Examples}
\label{sec:examples}

Our first example are step-like potentials. Namely, we assume that the potentials $V_\pm$  are constant: $V_-(x)\equiv -\b_-^2$ on $[-1,-0]$ and $V_+(x)\equiv -\b_+^2$ on $[0,1]$, while outside these segments the functions $V_\pm$ vanish identically. Here $\b_\pm \in \mathbb{C}\setminus\{0\}$.  It is easy to confirm  that condition (\ref{2.4}) is equivalent to   $\sin\b_- \sin\b_+ \ne0$.
The functions $X_\pm$ can be found explicitly, which leads to{
\begin{equation*}
F(k)=F_-(k)F_+(k)\b_-^{-2}\b_+^{-2},\qquad F_\pm(k)=2ik\sqrt{k^2+\b_\pm^2}\cot\sqrt{k^2+\b_\pm^2}+\b_\pm^2+2k^2.
\end{equation*}}
Then using the second series in {(\ref{2.11})}, we  find  a  three-terms approximation for   
$k_n$:
\begin{align*}
k_n\cong &\frac{\pi n}{2\ell} +\frac{\pi n}{4\ell^2} \left(\frac{\cot \b_+}{\b_+}+\frac{\cot \b_-}{\b_-}\right)
-\frac{1}{8\ell^3} \left(  \frac{ i\pi^2 n^2 }{\b_+^2\sin^2 \b_+} +\frac{i\pi^2 n^2}{\b_-^2\sin^2 \b_-}
-\pi n \left(\frac{\cot \b_+}{\b_+}+\frac{\cot\b_-}{\b_-}\right)^2
\right).
\end{align*}

\begin{figure}
	\centering
	\includegraphics[width=0.8\columnwidth]{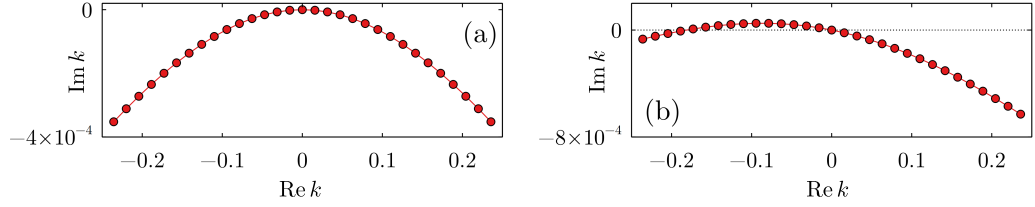}  
	\caption{(a) Sequence of wavenumbers corresponding to  resonances the self-adjoint step-function potential with  $\beta_+=1$ and $\beta_-=2i$. (b) Sequence of wavenumbers corresponding to  coexisting resonances (wavenumbers in the lower complex half-plane) and eigenvalues  (wavenumbers in the upper complex half-plane)} for the  non-self-adjoint potential with  $\beta_+=1$ and $\beta_-=2i+3$. For both panels, $\ell=100$.  The circles are $k_n$ and thin red lines are to guide the eye. 
	\label{fig:01}
\end{figure}

In order to provide an explicit illustration for the sequence of resonances and eigenvalues, we have considered several specific combinations of $\beta_+$ and $\beta_-$ and computed sequences $k_n$ using the iterative procedure (\ref{2.9}).  The number of iterations was chosen   to ensure
$|k_n-a_n - h^{[m]}_n(0)|< 10^{-16}$. For each combination of $\beta_+$ and $\beta_-$, values $r$ and $\max|F'|$ were   estimated   by plotting the graphs of   $F(k)$ and $F'(k)$ in the complex plane.  In  the examples shown in Fig.~\ref{fig:01} the prescribed accuracy   was   achieved after no more than 15 iterations.
The  case shown in Fig.~\ref{fig:01}(a) corresponds to the real-valued, i.e., self-adjoint potential. Respectively,  the  sequence consists only of resonances and  is situated in the  lower  complex half-plane of wavenumbers $k$. Figure~\ref{fig:01}(b) illustrates a non-self-adjoint case: here the sequence contains both resonances and eigenvalues.

Our second  example is a pair of two delta-interactions, namely, we assume that $V_\pm=\b_\pm \d(x)$, where $\b_\pm$ are   complex constants. Condition (\ref{2.4}) is ensured as $\b_\pm\ne0$.  Functions $X_\pm$ can be found explicitly and $F(k)$ reads  
\begin{equation}
	F(k)=(2ik-\b_+)(2ik-\b_-)\b_-^{-1}\b_+^{-1}.
\end{equation} 
The first three terms of the second series in (\ref{2.11}) are  
\begin{equation*}
k_n\cong  \frac{\pi n}{2\ell} - \frac{\pi n}{4\ell^2} \big(\b_+^{-1}+\b_-^{-1}\big)
-\frac{1}{8\ell^3} \left(i\pi^2 n^2  \big(\b_+^{-2}+\b_-^{-2}\big)
-\pi n \big(\b_+^{-1}+\b_-^{-1}\big)^2
\right).
\end{equation*} 
Numerical calculation  of $k_n$ for various $\b_\pm$ produces pictures similar to Fig.~\ref{fig:01}.

\section{Proof of Theorem~\ref{th1}}
\label{sec:proof}

The functions $X_-(0,k)$, $X_-'(0,k)$, $X_+(d_+,k)$, $X_+'(d_+,k)$ are entire in $k$, i.e., they are holomorphic in all $k\in\mathds{C}$. This can be
proved by reducing problem (\ref{2.3}) {for $X_\pm$} to the integral Volterra equations
\begin{equation*}
X_\pm(x,k)=e^{-ikx} + \int\limits_{-\infty}^{x} \frac{\sin k(x-t)}{k} V_\pm(t) X_\pm(t,k)\di t  
\end{equation*}
and studying them in a standard way.

As $k=0$, the denominator in the definition of $F$ is equal to $X_-'(0,0)X_+'(d_+,0)$ and by assumption is nonzero.
Hence, the function $F(k)$ is well-defined, holomorphic in the  circle $ B$ for some $r>0$. We also see that $F(0)=1$. Then we choose $r$ small enough so that the inequality holds:
\begin{equation}\label{3.4}
|F(k)-1|<1-e^{-\frac{\pi}{2}}\quad \text{as}\quad k\in\overline{ B}\qquad\Rightarrow\qquad e^{-\frac{\pi}{2}}<|F(k)|<2-e^{-\frac{\pi}{2}},\qquad k\in\overline{ B}.
\end{equation}

By $a_\pm(k)$, $b_\pm(k)$ we denote the   transmission and reflection coefficients for problems (\ref{2.3}), which are introduced as the coefficients in the identity
\begin{equation}\label{2.5}
X_\pm(x,k)=a_\pm(k)Y_\pm(x,-k) + b_\pm(k) Y_\pm(x,k).
\end{equation}
It is easy to confirm that
\begin{equation}\label{2.5a}
\begin{aligned}
&a_-(k)=\frac{ ik X_-(0,k)-X_-'(0,k)}{2ik},\qquad a_+(k)=\frac{ik X_+(d_+,k)-X_+'(d_+,k)}{2ik e^{-ikd_+}},
\\
&b_-(k)=\frac{ ik X_-(0,k)+X_-'(0,k)}{2ik},\qquad b_+(k)=\frac{ik X_+(d_+,k)+X_+'(d_+,k)}{2ik e^{ikd_+}}.
\end{aligned}
\end{equation}

We construct non-trivial solutions to problem (\ref{2.2}) as
$\psi(x,k)=X_-(x+\ell,{k})$, $x<\ell$.
Thanks to (\ref{2.3}), this function solves the equation in (\ref{2.2}) as $x<\ell$ and coincides with $e^{{i}k(x+\ell)}$ as $x<-\ell-d_-$. By (\ref{2.5}), as $-\ell<x<\ell$, the function $\psi$ reads
\begin{equation*}
\psi(x,k)={a_-(k)e^{-ik(x+\ell)} +b_-(k)e^{ik(x+\ell)}=a_-(k)e^{-2ik\ell}e^{-ik(x-\ell)} +b_-(k)e^{2ik\ell}e^{ik(x-\ell)}}.
\end{equation*}
The solution $\psi$ and its derivative $\psi'$ are to be continuous at $x=\ell$. Then we apply (\ref{2.5}) once again to obtain
\begin{equation*}
\psi(x,k)={a_-(k)e^{-2ik\ell} X_+(x-\ell,k) + b_-(k)e^{2ik\ell} X_-(x-\ell,-k)},\qquad x>\ell.
\end{equation*}
This  function solves the equation in (\ref{2.2}) as $x>\ell$, is $C^1$-smooth at $x=\ell$,  and it remains to find its asymptotic behavior as $x\to+\infty$. Again by (\ref{2.5}), for $x>\ell+d_+$ we get:
\begin{align*}
\psi(x,k)=&{\Big(a_-(k)a_+(k)e^{-2ik\ell} + b_-(k)b_+(-k)e^{2ik\ell}
\Big)e^{-ik(x-\ell)}}
\\
&+{\Big(a_-(k)b_+(k)e^{-2ik\ell} + b_-(k)a_-(-k)e^{2ik\ell}\Big)e^{ik(x-\ell)}}.
\end{align*}
Since the function $\psi$ should not involve $e^{{-}ikx}$ as $x\to+\infty$, the first term in the right hand side of the above formula should vanish. Expressing $a_\pm$, $b_\pm$ by formulae (\ref{2.5a}) and dividing the mentioned  term  by ${b_-(k)b_+(-k)}$, we arrive at the equation
\begin{equation}\label{3.3}
e^{4ik\ell}=F(k).
\end{equation}
The zeroes of this equation are values  $k$, for which problem (\ref{2.2}) possesses non-trivial solutions.

We proceed to proving (\ref{2.9}). Thanks to (\ref{3.4}), the function $F(k)$ ranges in a circle of radius $1-e^{-\frac{\pi}{2}}$ centered at $1$ and this circle does not contain the origin. Then the function $h(k):=\ln F(k)$ has no branching points as $k\in\overline{ B}$ and is holomorphic. We fix $n$ such that $|n|\leqslant N_\ell$. It is clear that $ B_n\subset B$ and we can rewrite equation (\ref{3.3}) as
\begin{equation}\label{3.7}
z=h_n(z+a_n),\qquad k=z+a_n,\qquad h_n=-\frac{i}{4\ell}h.
\end{equation}
Let us prove that the function   is a contracting mapping on $\overline{ B_0}$. It follows from (\ref{3.4}) that
\begin{equation}\label{3.13}
|h(k)|=|\ln F(k)|\leqslant \sum\limits_{m=1}^{\infty} \frac{|F(k)-1|^m}{m}\leqslant \sum\limits_{m=1}^{\infty} \frac{(1-e^{-\frac{\pi}{2}})^m}{m}=\frac{\pi}{2}
\end{equation}
and hence, $|h_n(k)|\leqslant \tfrac{\pi}{4\ell}$ as $k\in\overline{ B}$. This means that $h_n(z)\in \overline{ B_0}$ as $z\in\overline{ B}$. By the Lagrange formula and (\ref{3.4}), for all $z_1, z_2\in\overline{ B_0}$, we obtain:
\begin{equation}\label{3.9}
|h_n(z_1)-h_n(z_2)|=\left|\frac{i}{4\ell} \big(h(z_1+a_n)-h(z_2+a_n)\big)\right|= \frac{|F'(z_*+a_n)|}{4\ell|F(z_*+a_n)|}|z_1-z_2|\leqslant \frac{e^\frac{\pi}{2}\max\limits_{\overline{B}} |F'| }{4\ell}|z_1-z_2|,
\end{equation}
where $z_*$ is some point in the segment connecting the points $z_1$
and $z_2$. Hence, by (\ref{2.8}), the function $h_n(\cdot+a_n)$ is a contracting mapping. By the contracting mapping principle we obtain immediately that equation (\ref{3.7}) possesses the unique {solution $z_n$} in $\overline{B_0}$ and therefore, $k_n=z_n+a_n$ is the unique solution of (\ref{3.3}) in $\overline{B_n}$.  The solution $z_n$ can be represented as the limit of $h_n^{[m]}(0)$ as $m\to\infty$ and this proves representation for $k_n$ in
(\ref{2.9}). Applying (\ref{3.9}) and the estimate $|z_n|\leqslant \tfrac{\pi}{4\ell}$ to the identity
$z_n-h_n^{[m]}(0)=h_n^{[m]}(z_n)-h_n^{[m]}(0)$,
we arrive at the estimate in (\ref{2.9}).

Let us prove (\ref{2.11}), (\ref{2.12}), (\ref{2.12a}).  
By the standard estimate for the derivatives of a holomorphic function applied to the function $h$ and the circle $ B_n$ and by (\ref{3.13}) we get:
\begin{equation*}
\left|\frac{d^{m-1} h^m}{dk^{m-1}}(a_n)\right|\leqslant (m-1)! \left(\frac{4\ell}{\pi}\right)^{m-1} \max\limits_{\overline{ B}} |h^m|\leqslant \frac{\pi}{4\ell} (m-1)! (2\ell)^{m},\qquad
\left| \frac{(-i)^m}{4^m m!\ell^m} \frac{d^{m-1} h^m}{dk^{m-1}}(a_n)\right|\leqslant\frac{1}{2^m m}\frac{\pi}{4\ell},
\end{equation*}
and this proves the stated convergence of the first series in (\ref{2.11}). By $\tilde{z}_n$ we denote the sum of this series; the above estimates also imply that $|\tilde{z}_n-a_n|\leqslant \frac{\pi}{4\ell}\ln 2< \frac{\pi}{4\ell}$, $\tilde{z}_n\in B_n$.
Let us prove that $\tilde{z}_n$ solves equation (\ref{3.7}); due to the uniqueness of the root in $\overline{ B_0}$, this will imply $\tilde{z}_n=z_n$ and will prove the first identity in (\ref{2.11}).

The function $h_n$ is holomorphic and $\tilde{z}_n$ is holomorphic in $\ell$. Then we can write the Taylor expansion employing the Fa\`a~di~Bruno formula:
\begin{align*}
h_n(\tilde{z}_n+a_n)=&\sum\limits_{m=0}^{\infty} \frac{\tilde{z}_n^m}{m!} \frac{d^m h_n}{dk^m}(a_n)=\sum\limits_{m=0}^{\infty} \left(-\frac{i}{4\ell}\right)^{m+1}\frac{H_{m}^{(n)}(a_n)}{m!},
\qquad
H_m^{(n)}:=\sum\limits_{\Tht\in\Pi_m} \frac{d^{|\Tht|}h}{dk^{|\Tht|}}\prod\limits_{\tht\in\Tht}
\frac{d^{|\tht|-1}h^{|\tht|}}{dk^{|\tht|-1}},
\end{align*}
where $\Pi_m$ is the set of all (unordered) partitions of an $m$-element set,
the writing `$\tht\in\Tht$' means that $\tht$ runs through the list of all parts in a partition $\Tht$. By $|\Tht|$ and $|\tht|$ we denote the cardinalities of these sets. In view of the Leibnitz rule, we need to prove one of the following equivalent identities
\begin{equation}\label{3.15}
H_m^{(n)}=\frac{1}{m+1} \frac{d^m h^{m+1}}{dk^m} \quad \Leftrightarrow \quad
\frac{1}{m+1}(\underbrace{h+\ldots+h}\limits_{m+1\ \text{times}})^m=\sum\limits_{\Tht\in\Pi_m} h^{|\Tht|} \prod\limits_{\tht\in\Tht} (|\tht||h|)^{|\tht|-1}= \sum\limits_{\Tht\in\Pi_m} h^m \prod\limits_{\tht\in\Tht} (|\tht|)^{|\tht|-1}
\end{equation}
and it is sufficient to check the latter identity for $h=1$.
For each $p\in\mathds{N}$, the quantity $p^{p-1}$ is the
number of all (ordered) ways of splitting  a  $p$-element set into $p$ blocks including possible empty blocks. The set of all such splittings is denoted by $\hat{\Pi}_k$ and $p^{p-1}=\sum\limits_{\tau\in\hat{\Pi}_k} 1$. Then we can rewrite the right hand side in (\ref{3.15}) with $h=1$ as
\begin{equation}\label{3.16}
\sum\limits_{\Tht\in\Pi_m}  \prod\limits_{\tht\in\Tht} (|\tht|)^{|\tht|-1}=\sum\limits_{\Tht\in\Pi_m} \sum\limits_{\substack{\tau\in\hat{\Pi}_{|\tht|}\\
		\tht\in\Tht}}1=\sum\limits_{p=1}^{m}\sum\limits_{
	\Tht=\{\tht_1,\ldots,\tht_p\}\in\Pi_m} \sum\limits_{\tau_1\in\hat{\Pi}_{|\tht_1|}}\cdots
\sum\limits_{\tau_p\in\hat{\Pi}_{|\tht_p|}} 1.
\end{equation}
For each $\Tht=(\tht_1,\ldots,\tht_p)\in\Pi_m$, the sets $\tau_1\in\Pi_{\tht_1},\ldots,\tau_p\in\Pi_{\tht_p}$ provide one of possible splittings of an $(m-p)$-element set into $m$ blocks including possible empty blocks. These splittings are in one-to-one correspondence with the sets $\{\tau_1,1,\tau_2,2,\ldots,\tau_p,p\}$ being splittings of an $m$-element set. The total number of the latter sets is $\tfrac{(m-1)!}{(p-1)!(m-p)!}$, while the total number of ways of splitting an $(m-p)$-element set into $m$ blocks including possible empty blocks is $m^{m-p}$. Hence,
\begin{equation*}
\sum\limits_{
	\Tht=\{\tht_1,\ldots,\tht_p\}\in\Pi_m} \sum\limits_{\tau_1\in\hat{\Pi}_{|\tht_1|}}\cdots
\sum\limits_{\tau_p\in\hat{\Pi}_{|\tht_p|}} 1=\frac{(m-1)!}{(p-1)!(m-p)!}m^{m-p}.
\end{equation*}
Substituting this identities into (\ref{3.16}), we immediately arrive at (\ref{3.15}) and this proves the first identity in (\ref{2.11}).
For the coefficients in the first series in (\ref{2.11}) we have their Taylor expansions:
\begin{equation*}
\frac{d^{m-1}\ln^m F}{dk^{m-1}}(a_n)=\sum\limits_{p=0}^\infty \frac{1}{p!} \left(\frac{\pi n}{2\ell}\right)^p \frac{d^{m+p-1} \ln^m F}{dk^{m+p-1}}(0).
\end{equation*}
Substituting this formulae in the first series in (\ref{2.11}) and collecting the coefficients at the like powers of $\ell$, we arrive immediately at the second identity in (\ref{2.11}). The series in (\ref{2.11}) can be regarded as a Taylor series of $k_n$ at some point $a_n$ written in powers of $-\tfrac{i}{4\ell}$ and as a similar series at zero. Then estimates (\ref{2.12}), (\ref{2.12a}) are just   standard estimates for the remainder in the Lagrange form. The proof of Theorem~\ref{th1} is complete.

\section*{Acknowledgments}

We thank  a referee for useful remarks allowed us to improve the original version of the paper. The work of   D.A.Z. is supported  by Russian Foundation for Basic Research,  project No. 19-02-00193$\backslash$19.


\end{document}